# Simple models for dynamic hysteresis loop calculations: Application to hyperthermia optimization

J. Carrey*, B. Mehdaoui and M. Respaud

Université de Toulouse; INSA; UPS; LPCNO (Laboratoire de Physique et Chimie des Nano-Objets), 135 avenue de Rangueil, F-31077 Toulouse, France and
CNRS; UMR 5215; LPCNO, F-31077 Toulouse, France

**Abstract:**
To optimize the heating properties of magnetic nanoparticles (MNPs) in magnetic hyperthermia applications, it is necessary to calculate the area of their hysteresis loops in an alternating magnetic field. The three types of theories suitable for describing hysteresis loops of MNPs are presented and compared to numerical simulations: equilibrium functions, Stoner-Wohlfarth model based theories (SWMBTs) and a linear response theory (LRT) using the Néel-Brown relaxation time. The configuration where the easy axis of the MNPs are aligned with respect to the the magnetic field and the configuration of a random orientation of the easy axis are both studied. Suitable formulas to calculate the hysteresis areas of major cycles are deduced from SWMBTs and from numerical simulations; the domain of validity of the analytical formula is explicitly studied. In the case of minor cycles, the hysteresis area calculations are based on the LRT. A perfect agreement between the LRT and numerical simulations of hysteresis loops is obtained. The domain of validity of the LRT is explicitly studied. Formulas are proposed to calculate the hysteresis area at low field that are valid for any anisotropy of the MNP. The magnetic field dependence of the area is studied using numerical simulations: it follows power laws with a large range of exponents. Then, analytical expressions derived from the LRT and SWMBTs are used in their domains of validity for a theoretical study of magnetic hyperthermia. It is shown that LRT is only pertinent for MNPs with strong anisotropy and that SWMBTs should be used for weakly anisotropic MNPs. The optimum volume of MNPs for magnetic hyperthermia is derived as a function of material and experimental parameters. Formulas are proposed to allow to the calculation of the optimum volume for any anisotropy. The maximum achievable specific absorption rate (SAR) is calculated as a function of the MNP anisotropy. It is shown that an optimum anisotropy increases the SAR and reduces the detrimental effects of the size distribution of the MNPs. The optimum anisotropy is simple to calculate; it depends only on the magnetic field used in the hyperthermia experiments and the MNP magnetization. The theoretical optimum parameters are compared to those of several magnetic materials. A brief review of experimental results as well as a method to analyze them is proposed. This study helps in the determination of suitable and unsuitable materials for magnetic hyperthermia and provides accurate formulas to analyze experimental data. It is also aimed at providing a better understanding of magnetic hyperthermia to researchers working on this subject.

# Main Text:
## I. Introduction

Magnetic hyperthermia is a promising cancer treatment technique that is based on the fact that magnetic nanoparticles (MNPs) placed in an alternating magnetic field release heat. Active research is being done to improve the specific absorption rate (SAR) of MNPs, which could permit the treatment of tumors of a smaller size [1] and could reduce the amount of material that must be injected to treat a tumor of a given size.

If an assembly of MNPs is put into an alternating magnetic field of frequency $f$ and amplitude $\mu_0 H_{\max}$, the amount of heat $A$ released by the MNPs during one cycle of the magnetic field simply equals the area of their hysteresis loop, which can be expressed as

$$A = \int_{-H_{\max}}^{+H_{\max}} \mu_0 M(H) dH, \qquad (1)$$

where $M(H)$ is the NP magnetization. Then, the SAR is:

$$SAR = Af. \qquad (2)$$

As will be described in more detail below, $A$ depends, in a very complex manner, on the characteristics of the NPs: $A$ depends on the NPs' effective anisotropy $K_{\text{eff}}$, their volume $V$, the temperature $T$, the frequency and amplitude of the magnetic field and eventually magnetic interactions between NPs. It is thus crucial to be able to evaluate $A$ as precisely as possible as a function of these parameters in order to target the optimum parameters for the required application.

The theoretical literature on the properties of MNPs is very large. Of this large number of articles, those related to the evolution of the hysteresis area with the intrinsic parameters of the MNPs are of interest for magnetic hyperthermia applications [2, 3, 4, 5, 6, 7, 8]. A few theoretical papers have also been devoted specifically to the problem of magnetic hyperthermia [1, 9, 10, 11, 12, 13, 14]. However, the articles published on this subject are not complete and are sometimes inaccurate. First, the majority of the articles are mainly based on the linear response theory (LRT), which, as will be shown later, is not the most useful for magnetic hyperthermia. Second, when using theories derived from the Stoner-Wohlfarth model, the domain of validity is not taken into account and, more importantly, central conclusions that could be derived from them are missing. Third, in articles published by Hergt *et al.* an artificial separation of the mechanisms responsible for the heating is made between "hysteresis losses" and "relaxation losses". This separation is improper or at least very confusing; in our opinion, it is detrimental to a deep understanding of magnetic hyperthermia and to correct usage of the models to calculate SARs. Unfortunately, the paradigmatic presentation of magnetic hyperthermia by a large number of experimental articles still follows this separation. Finally, a recent article by N. A. Usov has used numerical simulations of hysteresis loops to study hyperthermia [14]. Although this article draws qualitatively correct conclusions, it is based on the study of examples and does not provide a generalization and a quantitative approach to the problem.

The present article aims to give a complete and rigorous presentation of the theory of magnetic hyperthermia. First, in Section II, we will provide a global view of the three types of theories suitable for the calculation of the hysteresis loop areas of MNPs, and we will give simple and accurate analytical formulas for this purpose; this will include a precise determination of their domain of validity using numerical calculations. A clarification of the issue concerning

"hysteresis losses" vs. "relaxation losses" will also be done in this section. Numerical simulations will be used to illustrate the variety of curves that could be obtained experimentally when measuring the magnetic field dependence of the SAR. In Section III, the previous results will be used for a specific study of magnetic hyperthermia. It will be shown that the LRT is only pertinent for strongly anisotropic MNPs; for weakly anisotropic MNPs, theories derived from the Stoner-Wohlfarth model should be used instead. Formulas that predict the optimum volume of MNPs as a function of material and experimental parameters will be provided. Additionally, it will be shown that the anisotropy of the MNPs is the central parameter for the optimization of magnetic hyperthermia because it determines both the maximum achievable SAR and controls the influence of the size distribution of MNPs on the SAR. A simple formula to determine the optimum anisotropy for magnetic hyperthermia will be proposed, and a comparison with the parameters of bulk magnetic materials will be done. This study should help to determine suitable and unsuitable materials for magnetic hyperthermia, and it will provide accurate formulas to analyze experimental data. We also hope it will lead to a better understanding of magnetic hyperthermia for researchers in this field.

## II. Numerical calculations and analytical expressions of hysteresis loops

### 1. Single-domain uniaxial nanoparticles in a magnetic field.

Let us consider a MNP of volume $V$ composed of a ferromagnetic material having a spontaneous magnetization $M_S$ and a magnetocrystalline anisotropy. Below a critical volume, the MNP becomes single-domain to minimize its magnetic energy. Because all the spins are parallel to one another, one can model the magnetization as a single giant magnetic moment $\mu = M_S V$, the amplitude of which does not depend on its spatial orientation; these are the so-called "macrospin" and coherent rotation approximations. As a result of magnetic anisotropy, $\mu$ is generally pinned along well-defined directions, i.e., along its magnetic anisotropy axis. As a result of several contributions, among them the magnetocrystalline, shape and surface contributions that arise from spherical deviations, the anisotropy can be very complex. Indeed, these numerous contributions have neither the same symmetries (cubic versus uniaxial) nor the same directions. Nevertheless, one of these contributions dominates and determines the main first-order contribution. From a practical point of view, one generally concludes from the experimental studies that the anisotropy displays a first-order-dominant uniaxial character. Thus, considering the macrospin approximation and an effective uniaxial anisotropy ($K_{eff}$), the energy of a MNP placed in an external magnetic field ($\mu_0 H_{max}$) is given by the following [2]:

$$E(\theta,\phi) = K_{eff} V \sin^2(\theta) - \mu_0 M_S V H_{max} \cos(\theta - \phi), \quad (3)$$

where $\theta$ is the angle between the easy axis and the magnetization and $\phi$ is the angle between the easy axis and the magnetic field [see Fig. 1(a)]. In the following, we will use the dimensionless parameters $\sigma = \dfrac{K_{eff} V}{k_B T}$ and $\xi = \dfrac{\mu_0 M_S V H_{max}}{k_B T}$. The reduced magnetic energy normalized to the thermal energy is

$$\frac{E(\theta,\phi)}{k_B T} = \sigma \sin^2(\theta) - \xi \cos(\theta - \phi). \quad (4)$$

Fig. 1(c) displays the reduced magnetic energy as a function of the normalized magnetic field ($\xi$ varies between 0 and 2) and the angle $\theta$; for a given particle orientation, $\phi = 30°$ and $\sigma = 1$. Two different shapes are noticeable. When $\mu_0 H_{max}$ is greater than the anisotropy field $\mu_0 H_K = 2K_{eff} /$ $M_S$, the energy landscape displays only one minimum, which defines the equilibrium position, i.e., along the anisotropy axis direction. Conversely, when $\mu_0 H_{max}$ is less than $\mu_0 H_K$, the energy profile as a function of $\theta$ displays two minima at the coordinates $(\theta_1, E_1)$ and $(\theta_2, E_2)$ and two maxima. We will refer to $(\theta_3, E_3)$ as the saddle point, i.e., the smaller maximum [see Fig. 1(b)]. For $\xi = 0$ (in the absence of magnetic fields), the magnetization can take two equivalent equilibrium values at $\theta_1 = 0°$ and $\theta_2 = 180°$, i.e., along its easy axis [see Figs. 1(c) and 1(d)]. For a finite positive $\xi$, the magnetic field favors one of the two minima (here, the one initially at $\theta_1 = 0°$). Increasing $\xi$ moves the abscissa of this minimum progressively so that a magnetization parallel to the magnetic field is favored. In the examples of Figs. 1(c) and 1(d), the abscissa of the minimum tends progressively toward $\theta_1 = \phi = 30°$. For a negative magnetic field, the energy landscape is similar to the positive one except that the well at $\theta_2 = 180°$ is initially favored and the minimum tends toward $\theta_2 = 210°$ for large $\xi$. Different shapes of the energy landscape can be more clearly seen in Fig. 1(d) where $E/k_B T$ is plotted versus $\theta$ for various values of $\xi$.

2. Equilibrium functions.

Let us first consider the case of thermal equilibrium, i.e., the derivation of the equilibrium functions characterized by a reversible magnetization process. Details on the calculation of the equilibrium functions can be found in [15]. The principle is the following: for MNPs whose easy axes are oriented at $\phi$ with respect to the magnetic field, the probability $f(\theta)$ to find the magnetization in a given direction is

$$f(\theta,\phi) = \frac{\exp\left(\frac{E(\theta,\phi)}{kT}\right)}{\int_\theta \exp\left(\frac{E(\theta,\phi)}{kT}\right)d\theta}. \qquad (5)$$

The resulting magnetization in the direction of the applied magnetic field is derived after numerical integrations [15]. The main results are displayed in Fig. 2 for two cases of interest: Fig. 2(a) displays the hysteresis loop with the hypothesis that all the anisotropy axes are oriented along the direction of the magnetic field ($\phi = 0$), and Fig. 2(b) illustrates the hypothesis that the measured system is an assembly of MNPs with their anisotropy axes randomly oriented in space. In the latter case, the hysteresis loop is then the result of an integration over all possible values for $\phi$. The magnetization $M$ is then given by

$$M = \int_0^{\pi/2} M(\phi)\sin\phi\, d\phi. \qquad (6)$$

Analytical expressions can be derived using Eqns. (3), (5) and (6) in two cases. First, if the anisotropy of the MNPs is neglected, i.e., for $\sigma = 0$, the magnetization reads

$$M = M_S L(\xi), \qquad (7)$$

where $L(\xi)$ is the Langevin function:

$$L(\xi) = \coth(\xi) - \frac{1}{\xi}. \qquad (8)$$

Thus, the widely used Langevin function intrinsically neglects the anisotropy of MNPs. The Langevin function is plotted in Figs. 2(a) and 2(b) for comparison with the numerical results. Second, when $\sigma$ is very large, one can consider that the magnetization has only two possible positions: the two minima of the energy landscape. This is called the "two-level approximation".

It reduces the integration over all possible values of $\theta$ to a sum of two terms. If one considers the case where the easy axis of the MNPs is aligned along the direction of the field ($\phi = 0$), the magnetization is expressed as

$$M = \frac{M_S \exp(\sigma+\xi) - M_S \exp(\sigma-\xi)}{\exp(\sigma+\xi) - \exp(\sigma-\xi)} = M_S \tanh(\xi). \qquad (9)$$

This function is also plotted for the sake of comparison with the numerical results in Fig. 2(a). It is observed for $\phi = 0$ that the magnetization curve progressively evolves from a $L(\xi)$ function for small $\sigma$ to the $\tanh(\xi)$ function for large $\sigma$.

Considering these general expressions, one can derive the equilibrium susceptibility in each case. To this end, Figs. 2(a) and 2(b) illustrate two results of interest: i) in all cases, the magnetization is linear with the magnetic field only when $\xi < 1$ and ii) for randomly oriented MNPs, the initial slope of the equilibrium function does not depend on the anisotropy of the MNPs and is the same as the one of the Langevin function. In this case, the static susceptibility is given by

$$\chi_{0\,Langevin} = \frac{\mu_0 M_S^2 V}{3 k_B T}. \qquad (10)$$

The case of MNPs with an easy axis aligned along the direction of the field is more complex: for $\sigma \ll 1$, the Langevin function is valid and leads to a susceptibility similar to Eqn. (10). However, for $\sigma \gg 1$ the equilibrium susceptibility $\chi_{0\,\tanh}$ of an aligned MNP is equal to the slope of the tanh function:

$$\chi_{0\,\tanh} = \frac{\mu_0 M_S^2 V}{k_B T} = 3 \chi_{0\,Langevin}. \qquad (\phi = 0) \qquad (11)$$

As a consequence, the equilibrium susceptibility of an assembly of MNPs with their easy axes aligned along the magnetic field evolves from $\chi_{0\,Langevin}$ for small $\sigma$ to $\chi_{0\,\tanh}$ for large $\sigma$. Fig. 2(c) displays this evolution as deduced from the numerical calculations of Fig. 2(a). A phenomenological fit of the $\chi_0(\sigma)$ function is provided and plotted in Fig. 2(c):

$$\frac{\chi_0(\sigma)}{\chi_{0\,Langevin}} = 3 - \frac{2}{1+\left(\dfrac{\sigma}{3.4}\right)^{1.47}}. \qquad (\phi = 0) \qquad (12)$$

3. Hysteresis loops at $T = 0$ – Stoner-Wohlfarth model.

Details on the calculations and results of the Stoner-Wohlfarth model can be found in several articles [2, 16]. The original Stoner-Wohlfarth model does not take into account any thermal activation, which is relevant at $T = 0$ or in the limit of infinite field frequency ($f \to \infty$). As a consequence of neglecting thermal activation, the magnetization can only stay along one of the two equilibrium positions. At $T = 0$ and when the magnetization is in one of the two minima of energy, the switch of the magnetization from the metastable state to the equilibrium position can only occurs when the energy barrier is fully removed by the magnetic field. The field at which this occurs is called the critical field [see Fig. 1(d)].

For a MNP with its easy axis aligned along the magnetic field direction, the suppression of the anisotropy barrier by the magnetic field occurs at $\mu_0 H_K$. The resulting hysteresis loop is a perfect square, i.e., the coercive field equals both the critical and the anisotropy field in this case. This is shown using numerical calculations in Fig. 3(a). The area of this hysteresis loop is maximal and is given as

$$A = 4\mu_0 H_C M_S = 4\mu_0 H_K M_S = 8 K_{eff}. \quad (13)$$

Note that upon tilting the angle $\phi$ of the MNP, the hysteresis loops progressively close up, and they become fully closed for $\phi = 90°$ [3] (not shown). As a consequence, the case of randomly oriented MNPs, exhibits a reduced coercive field $\mu_0 H_C = 0.48 \mu_0 H_K$ and a remnant magnetization that is half the saturation value because the hysteresis loop is an average over all contributions [see Fig. 3(b)]. The area of the hysteresis loop is reduced to

$$A = 2\mu_0 H_C M_S = 1.92 K_{eff}. \quad (14)$$

4. Dynamic hysteresis loop calculations within the two-level approximation.
a. Model

In between the pure superparamagnetic state, which is characterized by equilibrium functions where the magnetic moment can explore all directions, and the $T = 0$ limit, wherein the magnetic moment lies in its local minimum, the description of the field dependence of the magnetization is very complicated. Because the energy barrier of MNPs is relatively low, the magnetization reversal is thermally activated – the so-called Néel-Brown relaxation –, which leads to a progressive reduction of $\mu_0 H_C$ when the temperature is raised. Decreasing the sweeping rate of the magnetic field has similar consequences [3, 4, 5, 8]. The incorporation of these effects in a model is far from easy. Within the two–level approximation, one neglects excited states inside each well (which is strictly valid only if $\sigma \gg 1$) so that dynamic loop calculations only depend on the two minima $(\theta_1, E_1)$ and $(\theta_2, E_2)$ and on the saddle point $(\theta_3, E_3)$. When the applied magnetic field is below $\mu_0 H_K$, the magnetization can switch from the $\theta_1$ to the $\theta_2$ direction at a rate $\nu_1$ given by

$$\nu_1 = \nu_1^0 \exp\left(-\frac{E_3 - E_1}{kT}\right). \quad (15)$$

Similarly, the switching rate $\nu_2$ from the $\theta_2$ to the $\theta_1$ direction is given by

$$\nu_2 = \nu_2^0 \exp\left(-\frac{E_3 - E_2}{kT}\right). \quad (16)$$

The attempt frequencies $\nu_1^0$ and $\nu_2^0$ are complex functions of the material parameters (gyromagnetic ratio, damping, $M_S$ and $K_{eff}$) and experimental conditions (temperature and magnetic field) [17, 18]. For the sake of simplicity, we will keep these frequencies constant and equal to $10^{10}$ Hz.

In the various theoretical articles dealing with the influence of a finite temperature and frequency in the Stoner-Wohlfarth model, the numerical methods and approximations to include these thermally activated jumps vary. In their numerical simulations, Garcia-Otero *et al.* have taken the crude approximation that the switching occurs as soon as $\Delta E(\mu_0 H_{max}, \phi) = (E_3 - E_i) = k_B T$ where $i$ identifies the starting well [5]. Pfeiffer *et al.* assumed that the switch from one well to the other occurs when the relaxation time over the barrier matches a "measurement time" $\tau_m$ [4]. In both articles, the final results for the variation of the coercive field with temperature and frequency depends on this $\tau_m$ parameter. However, trying to define the value of $\tau_m$ has necessarily unphysical consequences. Indeed, the coercive field mainly depends on the *sweeping rate* of the magnetic field. In a SQUID measurement, one could simply take the "measurement time" as the time to measure one point, which corresponds to Pfeiffer *et al.*'s criteria. However, the coercive field would vary as a function of the step value. Alternatively, one could take the time of a complete cycle. In this case, the coercive field would vary with the maximum applied magnetic field.

The principle of our calculation is more rigorous and is similar to the one used by Lu *et al.* [3] and Usov *et al.* [8, 14]. A time-dependent magnetic field $H(t) = H_{max} \cos(\omega t)$ is applied to the MNP along a direction that makes an angle $\phi$ with respect to the easy axis. To compute the magnetization, one has to calculate the time dependence of $p_1$ and $p_2 = (1-p_1)$, the probability of finding the magnetization in the first and second potential wells, respectively. The time evolution of $p_1$ reads

$$\frac{\partial p_1}{\partial t} = (1 - p_1)\nu_2 - p_1\nu_1. \qquad (17)$$

Knowing the occupation probabilities, one can calculate the magnetization according to

$$M = M_S \left( p_1 \cos\theta_1 + (1 - p_1)\cos\theta_2 \right). \qquad (18)$$

The resolution of the time evolution of $p_1$ is performed using an explicit Runge-Kutta (2,3) method. With this method, the time step is not constant but becomes shorter when $p_1$ varies more, which ensures an optimum compromise between calculation time and precision. When there is only one minimum, $p_1$ is simply set equal to either zero or unity.

To calculate hysteresis loops for a random orientation of MNPs, 50 cycles with $\phi$ ranging from 0 to $\frac{\pi}{2}$ are calculated. Then, the magnetization is calculated according to Eqn. (6). We will show that using this single model, one can simulate major and minor hysteresis loops and the behavior in the framework of the LRT. For the latter cases, the initial conditions are set to $p_1 = p_2 = 0.5$, and several successive hysteresis loops are performed until the curve converges and becomes symmetrical with respect to the abscissa axis. Under most conditions, only 2 or 3 cycles are necessary to achieve convergence. Typical examples of the hysteresis loops generated will be shown in the following sections.

b. Temperature and frequency dependence of the coercive field

Historically, the first analytical expressions for the temperature dependence of the coercive field were based on the approximation of the "measurement time" previously described. In the $\phi = 0$ case, the expression of the coercive field reads [4]

$$\mu_0 H_C = \mu_0 H_K \left[ 1 - \left( \frac{k_B T}{KV} \left( \ln \frac{\tau_m}{\tau_0} \right) \right)^{\frac{1}{2}} \right], \qquad (\phi = 0) \qquad (19)$$

where $\tau_0$ is the frequency factor of the Néel-Brown relaxation time defined as (see Section II.5)

$$\tau_0 = \frac{1}{2v_1^0} = \frac{1}{2v_2^0}. \tag{20}$$

In the case of randomly oriented NPs, the following analytical expression was obtained by Garcia-Otero *et al.* [5]:

$$\mu_0 H_C \approx 0.48 \mu_0 H_K \left[1 - \left(\frac{k_B T}{KV}\left(\ln\frac{\tau_m}{\tau_0}\right)\right)^{\frac{3}{4}}\right]. \quad \text{(random orientation)} \tag{21}$$

We previously used the latter equation to interpret hyperthermia experiments with FeCo MNPs by stating that $\tau_m = \frac{1}{f}$ [19]. However, as mentioned above, these analytical formulas do not depend on the sweeping rate of the magnetic field $4H_{max}f$ but on this undefined $\tau_m$ parameter. Recently, Usov *et al.* [8] proposed a novel dimensionless parameter $\kappa$ for the variation of the coercive field that takes into account the sweeping rate. In the $\phi = 0$ case, the coercive field is

$$\mu_0 H_C = \mu_0 H_K \left(1 - \kappa^{\frac{1}{2}}\right) \qquad (\phi = 0) \tag{22}$$

with

$$\kappa = \frac{k_B T}{KV} \ln\left(\frac{k_B T}{4\mu_0 H_{max} M_S V f \tau_0}\right). \tag{23}$$

In Fig. 3(c), numerical calculations of hysteresis loops are compared to Eqn. (22). To achieve this, a large number of simulations were performed with parameters varying over a wide range of values: $f$ (10-400 kHz), $\mu_0 H_{max}$ (0.05-5 T), $K$ ($10^3$-$10^6$ J.m$^{-3}$), $T$ (0.5-500 K) and the spherical radii of the nanoparticles (1.5-30 nm). The normalized coercive field extracted from the hysteresis loop is then plotted as a function of $\kappa$. The fact that all the data fall onto a single master curve confirms the relevance of the dimensionless parameter proposed by Usov *et al.* [8]. Our simulations are in good agreement with the analytical expressions (22) and (23) derived by Usov *et al.* as long as $\kappa$ is below roughly 0.5.

For the random orientation case, Usov *et al.* derived an expression for the coercive field from the phenomenological fit of their numerical simulations. However, the authors made the assumption that the coercive field equals the critical field, which is not rigorously true for NPs with a large $\phi$, and performed a fit over a large range of temperatures. In Fig. 3(d), our simulations for the coercive field in the random orientation case are shown. From the best fit of these data at low values of $\kappa$, the following formula is obtained:

$$\mu_0 H_C = 0.48 \mu_0 H_K \left(b - \kappa^n\right), \quad \text{(random orientation)} \tag{24}$$

where $b = 1$ and $n = 0.8 \pm 0.05$. Usov *et al.* found slightly different coefficients of $b = 0.9$ and $n = 1$. The domain of validity is roughly the same as for the aligned case : Eqn. (24) is roughly valid up to $\kappa = 0.5$. In the remainder of this article, our own values for $b$ and $n$ will be used.

   c. <u>Temperature and frequency dependence of the hysteresis loop area.</u>
   The aim of this subsection is to study the frequency and temperature dependencies of the hysteresis area and to derive general analytical expressions in the case of aligned and randomly oriented MNPs. This topic has not been addressed in the publications mentioned above. At $T = 0$, the area is proportional to the coercive field as given by Eqns. (13) and (14), and the question is whether these expressions are still valid when $T \neq 0$. Similar to the study of the coercive field dependence, we estimated from numerical hysteresis calculations the associated area for the aligned and randomly oriented cases; these are displayed in Figs. 3(e) and 3(f), respectively. Examples of these hysteresis loops are shown in Figs. 3(a) and 3(b). These data are then compared to those calculated using the analytical expressions

$$A(T) \approx 4\mu_0 H_C(T) M_S \qquad (\phi = 0) \qquad (25)$$

and

$$A(T) \approx 2\mu_0 H_C(T) M_S. \qquad \text{(random orientation)} \qquad (26)$$

The area $A(T)$ has been calculated by using i) $H_C$ deduced from the simulated hysteresis loops (shown as a dashed line) and ii) $H_C$ calculated using Eqns. (22) and (24) (shown as a solid line). The $A(T)$ curves calculated according to the first procedure match the exact values of the area except at large $\kappa$ values (above 1). This difference is due to the reduced squareness of the hysteresis loops. From the close comparison between the dots and the dashed line, it can be observed that Eqns. (25) and (26) slightly overestimate the hysteresis area for large $\kappa$. This overestimation partially compensates for the underestimation of the coercive field by Eqns. (22) and (24) at high $\kappa$. As a consequence, the combination of Eqns. (22) and (24) with Eqns. (25) and (26) gives an acceptable value of the area at higher $\kappa$ values than Eqns. (22) and (24) do for the coercive field. To provide a numerical limit that can be used later in the article, Figs. 3(e) and 3(f) show that the area is calculated with less than 10% error when $\kappa < 0.7$.

Finally, the transition toward reversible hysteresis loops can also be deduced from this figure. If we state that the reversibility occurs when $H_C \approx 0.01 H_K$, this transition can be estimated to occur when $\kappa \approx 1.6$.

### 5. Minor hysteresis loops and linear response theory.

The LRT has been previously reported in several articles [1, 7, 9, 10]. The presentation here will be slightly different from that of other articles and will aim to explicitly illustrate the fact that the LRT is also a model to calculate the hysteresis area, a point that was not always developed in previous works. The results for MNPs aligned with the magnetic field, not derived in previous articles, will also be given.

The LRT is a model that aims to describe the dynamic response of an assembly of MNPs using the Néel-Brown relaxation time. The starting assumption of this model is that the magnetic system responds linearly with the magnetic field and its magnetization can be put in the form

$$M(t) = \tilde{\chi} H(t), \qquad (27)$$

where $\tilde{\chi}$ is the complex susceptibility and reads

$$\tilde{\chi} = \chi_0 \frac{1}{1 + i\omega \tau_R}. \qquad (28)$$

$\chi_0$ is the static susceptibility defined in Section II.2, and $\tau_R$ is the time it takes for the system to relax back to equilibrium after a small step in the magnetic field. The results of Section II.2 showed that the magnetization is linear with the magnetic field approximately for the condition when $\xi < 1$. A small value of $\xi$ is thus the first criterion for the validity of the LRT; it will be more precisely studied below. Moreover, in Eqn. (28), $\tau_R$ is a variable independent of the magnetic field, which is only true for small deformations of the barrier between the two equilibrium positions, i.e., when $\mu_0 H_{max} \ll H_K$. At the magnetic field frequencies used in hyperthermia or magnetic measurements, it can be shown that the second condition is always verified when the first one is (see below). The relaxation time of the magnetization $\tau_R$ when the MNPs cannot move physically equals the Néel-Brown relaxation time $\tau_N$, which reads

$$\tau_R = \tau_N = \frac{1}{2\nu_1^0}\exp\left(\frac{K_{eff}V}{k_BT}\right) = \tau_0 \exp\left(\frac{K_{eff}V}{k_BT}\right). \qquad (29)$$

The fact that there is a factor $1/2$ between $\tau_0$ and the attempt frequency $\nu_1^0$ comes from the fact that we are dealing with a reversible jump in a system with two potential wells. If this point is unclear to the reader, it is illustrated in Fig. 4. Fig. 4 displays an imaginary case in which all of the MNPs are first magnetized in one direction and then relax at zero magnetic field through a reversible jump over an energy barrier at a rate $\nu_1$. The probability $p_1$ to find the MNPs in this well drops exponentially to 0.5 with a time constant of $\frac{1}{2\nu_1}$. Thus, the relaxation time of the magnetization is half the mean time taken by the magnetization to reverse spontaneously. This explains the factor of $1/2$ between $\tau_0$ and the attempt frequency $\nu_1^0$.

In the LRT, the response of the system to an alternating magnetic field
$$H(t) = H_{max}\cos(\omega t) \qquad (30)$$
is
$$M(t) = |\chi|H_{max}\cos(\omega t + \varphi), \qquad (31)$$
where $\varphi$ is the phase delay between the magnetization and the magnetic field. From Eqn. (28), it is straightforward to show that

$$|\chi| = \frac{\chi_0}{\sqrt{1+\omega^2\tau_R^2}} \qquad (32)$$

and

$$\sin\varphi = \frac{\omega\tau_R}{\sqrt{1+\omega^2\tau_R^2}} \quad \text{or} \quad \cos\varphi = \frac{1}{\sqrt{1+\omega^2\tau_R^2}}. \qquad (33)$$

Basic mathematics indicates that Eqns. (30) and (31) correspond to the parametric equation of an ellipse in the $(H, M)$ plane. The area $A_{ellipse}$ of this ellipse and the angle $\gamma$ between its long axis and the abscise axis are given by

$$A_{ellipse} = \pi H_{max}^2|\chi|\sin\varphi = \pi H_{max}^2\chi_0\frac{\omega\tau_R}{1+\omega^2\tau_R^2} \qquad (34)$$

and

$$\tan 2\gamma = \frac{2H_{max}^2|\chi|\cos\varphi}{H_{max}^2 - H_{max}^2|\chi|^2} = \frac{2\chi_0}{1+\omega^2\tau_R^2 - \chi_0^2}. \qquad (35)$$

In Fig. 5, the results of these equations are shown: Fig. 5(a) displays the ellipses plotted using Eqns. (30) and (31) for $\chi_0 = 0.1$ and $H_{max} = 1$ for various values of $\omega\tau_R$ while Fig. 5(b) displays the evolution of $\gamma$ from Eqn. (35) and from a graphical analysis of Fig. 5(a). A similar agreement is found when plotting the evolution of the hysteresis area $A$ using Eqn. (34) and by integrating over the area of the hysteresis loops of Fig. 5(a) (not shown). These figures and the corresponding equations illustrate the behavior of the magnetization loops as a function of the applied magnetic field in the LRT. First, when $\omega\tau_R \to 0$, the hysteresis loop is simply a straight line with a null hysteresis. In this condition, the angle $\gamma$ is such that $\tan\gamma(\omega\tau_R = 0) = \chi_0$. Then, the hysteresis area is maximal for $\omega\tau_R = 1$. When $\omega\tau_R \to \infty$, the system does not have the time to respond to the magnetic field excitation and $|\chi| \to 0$ as does the hysteresis area [7].

For magnetic hyperthermia, calculating the hysteresis area when a magnetic field $\mu_0 H_{max}$ is applied requires a combination of the $\chi_0$ expressions given by Eqns. (10), (11) or (12) with Eqn. (34); the final result is multiplied by $\mu_0$ [see Eqn. (1)]. For randomly oriented MNPs or for aligned MNPs when $\sigma$ is negligible, this leads to

$$A = \frac{\pi\mu_0^2 H_{max}^2 M_S^2 V}{3k_B T} \frac{\omega\tau_R}{(1+\omega^2\tau_R^2)}. \qquad \text{(random orientation) or } (\phi = 0 \text{ and } \sigma \ll 1) \quad (36)$$

For aligned MNPs with a strong $\sigma$, this leads to

$$A = \frac{\pi\mu_0^2 H_{max}^2 M_S^2 V}{k_B T} \frac{\omega\tau_R}{(1+\omega^2\tau_R^2)}. \qquad (\phi = 0 \text{ and } \sigma \gg 1) \quad (37)$$

For aligned MNPs with any $\sigma$, the phenomenological law given by Eqn. (12) can be used, which leads to

$$A = \frac{\pi\mu_0^2 H_{max}^2 M_S^2 V}{3k_B T} \frac{\omega\tau_R}{(1+\omega^2\tau_R^2)} \left(3 - \frac{2}{1+\left(\frac{\sigma}{3.4}\right)^{1.47}}\right). \qquad (\phi = 0 \text{ and any } \sigma) \quad (38)$$

In several articles, Eqn. (36) is defined as applying to "relaxation losses" of superparamagnetic MNPs. In these articles, "relaxation losses" are opposed − as if it was a different process − to the "hysteresis losses" of ferromagnetic NPs [1, 9, 12, 13]. We have tried to illustrate here that this distinction is not correct, or is at least confusing: all the losses, whether the MNPs are in the superparamagnetic regime or in the ferromagnetic regime, are always "hysteresis losses" insofar as they are simply given by the hysteresis loop area. LRT is simply one model among several that aims to calculate the hysteresis loop area and shape when the magnetic response is linear with the applied magnetic field. In order to avoid confusion and misunderstandings of this concept, we suggest putting an end to the distinction between *hysteresis losses* and *relaxation losses* and, rather, making a distinction between different kinds of *models* aiming at calculating the hysteresis area. For instance, it is correct to say "LRT is suitable to calculate the hysteresis area of MNPs in the superparamagnetic regime at low magnetic field" but not to say "in ferromagnetic NPs, relaxation losses disappear and are replaced by hysteresis losses". Unfortunately, misconceptions similar to this are present in a large number of articles on magnetic hyperthermia.

Numerical simulations have been performed to check the validity of Eqns. (29), (36) and (37). Specifically, numerical simulations of minor hysteresis loops were run with $K_{eff}$, $V$, $f$ and $\mu_0 H_{max}$ varying over a wide range of values while keeping $\xi \ll 1$ and $H_{max}/H_K \ll 1$. Hysteresis areas are then normalized by the prefactor of Eqns. (36) and (37) and plotted as a function of $\omega\tau_R$. The final results for $\phi = 0$ and for the random orientation case are plotted in Figs. 5(c) and 5(d). The fact that the hysteresis area displays a maximum for $\omega\tau_R = 1$ explains the shape of the curves. These graphs illustrate the perfect agreement between simulations and LRT both for the $\phi$

= 0 and the random orientation case. As a matter of fact, the hysteresis loops obtained by the numerical simulations are indistinguishable from the ones obtained using Eqns. (30), (31), (32) and (33), and so an illustration of the hysteresis loops obtained by numerical simulations would be indistinguishable from Fig. 5(a).

Next, the domain of validity of the LRT was studied by increasing $\xi$ and $H_{max} / H_K$ and comparing the hysteresis areas provided by simulations to those provided by Eqns. (36) and (37). The results are shown in Figs. 5(e) and 5(f). Practically, this study is performed by studying the volume dependence of the hysteresis area (which modifies $\xi$ only) for various values of $K_{eff}$ (which modifies $H_{max} / H_K$ only). It must first be noted that for realistic values of the measurement frequency, $\xi$ and $H_{max} / H_K$ are not completely independent. Indeed, combining Eqn. (39) with the definition of $\xi$ and $H_{max} / H_K$ shows that near the resonance – when the hysteresis area is not too weak - $\xi$ is always larger than $H_{max} / H_K$ by a factor of $2\ln(\omega\tau_0)$, which is always much larger than 1. As a consequence, there is no realistic case where $\xi<<1$, $H_{max} / H_K \approx 1$ and the area is not negligibly small.

As expected, increasing $\xi$ and $H_{max} / H_K$ leads to a discrepancy between the equations and simulations. When decreasing the anisotropy, the position of the peak in the area obtained by the simulations is progressively shifted toward higher values of $\xi$ compared to the position of the peak calculated using LRT. The discrepancy is greater in the random orientation case than the $\phi = 0$ case. To obtain quantitative values of the error made when using LRT, the ratio between the area given by simulations and calculations has been plotted in Figs. 5(g) and 5(h). In these graphs, the corresponding values of $H_{max} / H_K$ are provided. For $\xi = 1$, the discrepancy is around ±20 % for the $\phi = 0$ case and around +70%/-40% for the random orientation case. If lower error bars are required when using LRT, the maximum acceptable $\xi$ value should be reduced accordingly. Interestingly, it is observed that the LRT either overestimates or underestimates the area and that the transition between the two zones is approximately localized around the peak in the area [see Figs. 5(e) and 5(f)]. Therefore, if we identify the zone to the left of the peak as a "superparamagnetic regime" and the zone to the right of the peak as a "ferromagnetic regime", these data can be summarized this way: for values of $\xi$ above 1, the LRT overestimates the hysteresis area in the superparamagnetic regime and underestimates it in the ferromagnetic regime. Another conclusion, which will be developed later in this article but is also visible in Figs. 5(e) and 5(f), is that LRT is mainly useful for highly anisotropic NPs.

6. Dynamic hysteresis loops and area in the general case.

We have just seen that the LRT allows one to calculate the hysteresis area when $\xi < 1$. Similarly, SWMBTs can be used when $\kappa < 0.7$ and when the hysteresis loop is a major hysteresis loop, i.e., when the NPs are saturated by the magnetic field. In all other cases, these theories cannot be used, and numerical simulations are the only way to calculate the hysteresis area. In this subsection, we will present results for the hysteresis area provided completely by numerical simulations. In particular, numerical simulations give us the opportunity to study the magnetic field dependence of the hysteresis area, which is accessible in hyperthermia experiments by performing measurements as a function of the magnetic field. Because the results depend on all the external and structural parameters, there is no universal curve or pertinent dimensionless parameters. Thus, we have only used as an illustration the magnetic parameters of bulk magnetite and external parameters typical of hyperthermia. When they are not being varied, the parameter values in this section are $K_{eff}$ = 13000 J.m$^{-3}$, $M_S$ = 10$^6$ A.m$^{-1}$, $f$ = 100 kHz, $\mu_0 H_{max}$ = 20 mT, $T$ = 300 K and $\nu_0^1$ =10$^{10}$ Hz.

In Fig. 6(a), the hysteresis area is plotted as a function of the radius and temperature and in Fig. 6(b) as a function of radius and magnetic field in the $\phi = 0$ case. In both cases, it is evident that the largest areas are obtained for large ferromagnetic nanoparticles. However, for such nanoparticles abrupt transitions are observed as a function of the temperature or the applied magnetic field between a regime where the area is very small and a regime where the area is very large. Basically, when the coercive field is larger (smaller) than the applied magnetic field, the area is very small (very large). It is also observed that for a given set of parameters, there is an optimum radius to maximize the area, which we have plotted in the two graphs. The analytical determinations of this optimum volume and area will be the subject of Section III.

In Fig. 6(c) and 6(d), the magnetic field dependence of the hysteresis area is plotted for values of $\mu_0 H_{max}$ between 0 and $H_K$ (which here is 26 mT) in the $\phi = 0$ case and the random orientation case; the area is normalized to its value at 26 mT. For very small NPs, the LRT is valid and predicts a square dependence of the hysteresis loop area [see Eqns. (36)-(38)], which has been verified in a large number of experimental works (e.g., Refs 20, 21 or 22) and is also observed here. For large NPs in the ferromagnetic regime, the magnetic field dependence displays a very abrupt jump with a null hysteresis area below the critical fields and a sharp increase followed by a plateau. For NPs with intermediate sizes, the transition between these two regimes is progressive and the curves display a large variety of shapes. These curves were fitted by a power law in a range where the fit is acceptable. The corresponding exponents are shown in Fig. 7.

We will describe in detail the results for the $\phi = 0$ case [see Fig. 6(c) and Fig. 7], with the understanding that the random orientation case is qualitatively similar. The exponent of 2 predicted by the LRT is always observed when $\xi < 1$. For very small MNPs in the superparamagnetic regime (here, 3 nm), this square law is followed across the whole range of magnetic fields studied. For larger MNPs, the domain of validity of the LRT very rapidly shrinks, and this exponent is still observed at very small magnetic fields. However, the general shape of the curve for MNPs between 3.5 and 9 nm for magnetic fields up to 26 mT is a power law function with an exponent progressively decreasing from 2 down to 0.6. The fit by the power law is good over the whole range of magnetic fields studied. Above 9 nm, the MNPs are in the ferromagnetic regime where the curves display an inflexion point and an abrupt increase at the coercive field. In this case, the curves were fitted by a power law only up to this inflexion point. The exponent of the power law rises very quickly up to very large values as MNPs grow in size. As a consequence, a power law with a large range of exponents can be observed experimentally in the magnetic field dependence of SAR even in the simplest case of monodisperse single-domain nanoparticles. Thus, exponents other than 2 should not be considered as something exotic in hyperthermia experiments, and an exponent of 3 should not necessarily be considered as the signature of multi-domain nanoparticles.

7. <u>Summary of the models and magnetic properties as a function of the NP size</u>

In this subsection, we briefly summarize graphically the results of the previous sections to illustrate which model should be used to calculate the hysteresis area. This part is based on the description of Fig. 8, starting from the properties of small MNPs.

For small nanoparticles (when approximately $\kappa > 1.6$), the hysteresis loop is reversible, and the coercive field is almost null. In this case, the hysteresis loops can be calculated using equilibrium functions for any value of the magnetic field (see Section II.2). NPs in this range are useless for magnetic hyperthermia because of their null hysteresis, but the equilibrium functions are useful for two reasons: i) they are used in the LRT to calculate the initial static susceptibility $\chi_0$ and ii) they are useful to accurately fit magnetic measurements on MNPs in the superparamagnetic regime [15].

When the volume increases such that $\kappa < 1.6$, the hysteresis loop progressively opens. In this case, the shape of the hysteresis loop cannot be calculated simply for any value of $\mu_0 H_{max}$. However, the LRT allows one to calculate the hysteresis loop shape if $\xi < 1$ or $\xi \ll 1$ depending on the accuracy required (see Section II.5). The fact that the magnetic field value for which the LRT is valid progressively reduces as the NP volume increases is schematized at Label (1) in Fig. 8.

The formal transition between the superparamagnetic regime ($\omega \tau_N < 1$) and the ferromagnetic regime ($\omega \tau_N > 1$) occurs at $\omega \tau_N = 1$. Precisely at this transition, the hysteresis loop area *for small magnetic fields* displays a maximum (see Section II.5). However, we emphasize that nothing special occurs at this transition with respect to the hysteresis loop shape and area at high magnetic field: the coercive field has started to grow well before the transition and keeps increasing after the transition. This means that the hysteresis loop area at high magnetic fields does not display a maximum here but continues to increase with an increase in the volume.

For $\omega \tau_N > 1$, the MNPs are in the ferromagnetic regime where they display a more and more open hysteresis loop as their volume increases. In the ferromagnetic regime, the SWMBTs are suitable to describe the NP hysteresis loops if the NPs are not too close to the superparamagnetic-ferromagnetic transition, i.e., for $\kappa < 0.7$. Using SWMBTs to calculate the area supposes also that the MNPs are saturated, which is true for approximately $\mu_0 H_{max} > \mu_0 H_C$ in the $\phi = 0$ case and $\mu_0 H_{max} > 2\mu_0 H_C$ in the random orientation case. The LRT is still valid in this region and can be used to calculate minor hysteresis loop area at very low fields.

In larger MNPs, incoherent reversal modes start to occur, which lead to a decrease of the coercive field. This is where the theories used in this article cease to be valid. For large MNPs, SWMBTs predict for the coercive field a value independent of the NP volume, which is the value of the coercive field at $T = 0$. As a consequence, if the volume at which this phenomenon occurs is smaller than the one at which incoherent reversal modes start, a plateau in the evolution of the coercive field with the volume might in principle be observed. We made this assumption in Fig. 8, and the plateau is labeled as (3). If incoherent reversal modes began before this plateau, a peak in the coercive field value should be observed instead of a plateau. Finally, the largest MNPs are composed of a vortex [23] or of several magnetic domains separated by magnetic walls. In the latter case, the process leading to their magnetization is the growth of one or several domains in the direction of the field at the expense of the others. In this case, their hysteresis loops at very small magnetic fields are described by "Rayleigh loops" [24].

### III. <u>Optimum parameters for magnetic hyperthermia</u>

In this section, the models presented above are used to calculate the optimum parameters of MNPs for magnetic hyperthermia. The domain of validity of each model will be taken into account. The case of MNPs aligned with the magnetic field ($\phi = 0$) as well as the case of a random orientation will both be treated.

1. Optimum size as a function of the anisotropy

In this part, the optimum volume of MNPs for magnetic hyperthermia will be calculated as a function of their anisotropy. In the figures, specific values of the external parameters have been used: $f = 100$ kHz, $\mu_0 H_{max} = 20$ mT, $T = 300$ K and $\nu_0^1 = \nu_0^2 = 10^{10}$ Hz. These $f$ and $\mu_0 H_{max}$ values are the ones used in clinical applications at the Charité Hospital, Berlin [25]. In addition, results for three different values of $M_S$ ($M_1 = 0.4 \times 10^6$ A.m$^{-1}$, $M_2 = 10^6$ A.m$^{-1}$ and $M_3 = 1.7 \times 10^6$ A.m$^{-1}$) will be shown. They correspond to the magnetizations of $CoFe_2O_4$, magnetite and iron, respectively. Equivalent graphs for any value of the external and magnetic parameters can be plotted with the condition to resolve one equation numerically (see below).

In the LRT, even though the hysteresis area is different for $\phi = 0$ and for a random orientation of NPs, Eqns. (36), (37) and (38) show that the maximum of this area always occurs for $\omega \tau_N = 1$. This means that the optimum volume is given by the following equation, which is plotted in the two graphs of Fig. 9:

$$V_{opt} = \frac{K_B T}{K_{eff}} \ln(\pi f \tau_0). \qquad (39)$$

Because the LRT is valid when $\xi < 1$, this condition is plotted in the graphs of Fig. 9: it appears as horizontal lines above which the LRT is no longer valid. It is evident from this graph that the LRT is mainly useful for strongly anisotropic NPs. For instance, it is deduced from the intersection between Eqn. (39) and the function $\xi = 1$ that for $M_S = 10^6$ Am$^{-1}$, Eqn. (39) is no longer valid for MNPs with an anisotropy below $2 \times 10^5$ Jm$^{-3}$.

We now consider the optimum sizes predicted by the SWMDTs starting with the formula derived by Pfeiffer *et al.* and Garcia-Otero *et al.* [Eqns. (19) and (21)]. First, $\tau_m$ must be replaced by an expression depending on experimental parameters. For reasons that will be clear later, we arbitrarily state $\tau_m = \frac{1}{2\pi f}$. Then, it should be decided what the optimum coercive field of the MNPs is when a given magnetic field is applied. In the case of MNPs aligned with the field ($\phi = 0$) and due to the fact that the hysteresis loop is approximately square,

$$\mu_0 H_C \approx \mu_0 H_{max} \qquad (\phi = 0) \qquad (40)$$

is taken. This leads to an optimum area $A_{opt}$ given by

$$A_{opt} \approx 4\mu_0 M_S H_{max}. \qquad (\phi = 0) \qquad (41)$$

In the random orientation case, the magnetic field necessary to saturate an assembly of MNPs is approximately twice its coercive field. As a consequence, a coercive field of half the applied magnetic field could be targeted. Eqn. (26) shows this would lead to an optimum area $A_{opt} = \mu_0 M_S H_{max}$. However, numerical calculations show that it is better to target a coercive field slightly higher than this: the increase in area due to the increase in coercive field compensates for the fact that some of the MNPs are not switched by the applied magnetic field. The best compromise is found to depend slightly on the exact shape of the hysteresis loop and is given by

$$\mu_0 H_C \approx 0.81 \pm 0.04 \mu_0 H_{max}. \qquad \text{(random orientation)} \qquad (42)$$

With this optimum coercive field, the optimum area is

$$A_{opt} \approx 1.56 \pm 0.08 \mu_0 M_S H_{max}. \text{ (random orientation)} \qquad (43)$$

Combining Eqns. (40) and (42) with Eqns. (19) and (21) allows one to calculate the optimum volume $V_{opt}$. For NPs aligned with the magnetic field, this leads to

$$V_{opt} = \frac{-k_B T \ln(\pi f \tau_0)}{K\left(1 - \frac{\mu_0 H_{max} M_S}{2K}\right)^2}. \quad (\phi = 0) \qquad (44)$$

In the random case, this leads to

$$V_{opt} = \frac{-k_B T \ln(\pi f \tau_0)}{K\left(1 - \frac{1.69 \mu_0 H_{max} M_S}{2K}\right)^{\frac{4}{3}}}. \qquad \text{(random orientation)} \qquad (45)$$

These two functions are plotted in Figs. 9(a) and 9(b) in dashed lines.

Eqns. (22) and (24) are more rigorous methods to calculate the coercive field and thus the optimum volume. In these equations, a numerical solution is required to extract the volume corresponding to a given coercive field, the result of which is plotted in Figs. 9(a) and 9(b) in solid lines. The difference between the results provided by Eqns. (44) and (45) and the numerical solution of Eqns. (22) and (24) can reach up to 3 nm for the set of parameters we used, which is not negligible.

Figs 9(a) and 9(b) give evidence that the optimum volume obtained using SWMBTs deviates from the LRT results for small anisotropies, i.e., precisely in the domain where the LRT is not valid anymore. Thus, in this domain SWMBTs should be used instead of LRT to calculate the optimum size. For strong anisotropies, the optimum volumes given by Eqns. (44) and (45) tend toward the one deduced from LRT because of the assumption we made; i.e., that $\tau_m = \frac{1}{2\pi f}$.

The numerical solutions of Eqns. (22) and (24) also leads to an optimum volume very close to the one predicted by the LRT for strong anisotropy; the difference between the two predictions never exceeds 1 nm over a wide range of parameters: $f$ (5-500 kHz), $M_S$ (0.4-1.7×10$^6$ A.m$^{-1}$), $\tau_0$ (10$^{-9}$-10$^{-12}$ s) and $\mu_0 H_{max}$ (0-80 mT). Strictly speaking, there is a zone where none of the models used is valid because just above the LRT limit (when $\xi > 1$), $\kappa$ is not immediately smaller than 0.7. This point will be more clearly evidenced in the next section. However, because the SWMBT results approximately tend toward the LRT results at high anisotropy, it can be reasonably assumed that the transition between the two models is also acceptably reproduced by SWMBTs.

As a conclusion, Eqns. (44) and (45) can be used to give an approximate value of the optimum size of MNPs, but the error for weakly anisotropic MNPs can be significant. The numerical solution of Eqns. (22) and (24) can safely be used to calculate the optimum size of MNPs for magnetic hyperthermia over a wide range of anisotropies without caring too much about which is the most suitable model to describe their behavior. However, the most rigorous approach consists of calculating $\xi$ and $\kappa$ and using LRT when $\xi < 1$, Eqns. (22) and (24) when $\kappa < 0.7$ and numerical simulations otherwise.

### 2. Optimum anisotropy

In the previous section, the optimum size for MNPs with a given anisotropy was derived. However, the question of whether there is an optimum anisotropy was not addressed; this important point is treated now. To solve this problem, the SAR of a MNP with an optimum size was calculated versus the anisotropy. The calculations were performed for $M_S = M_2 = 10^6$ Am$^{-1}$ (magnetite value) and with the same values as given previously for the other parameters. To express the result in W/g, which is the usual unity for SARs, a density of $\rho = 5.2 \times 10^6$ kg.m$^{-3}$ (magnetite value) is assumed. The results of the LRT for the $\phi = 0$ and random orientation case were obtained using Eqns. (36), (38) and (39). They are plotted in Fig. 10 as solid lines. The data are not plotted outside the domain of validity of the LRT, i.e., when $\xi > 1$. The calculations in the ferromagnetic regime were performed for simplicity without numerical solutions using Eqns. (19), (21) and (40)-(45). This simplification does not change the main conclusions of this part. Data are plotted in Fig. 10 as solid lines and are not plotted outside the domain of validity of the SWMBTs, i.e., when $\kappa > 0.7$.

A very important point to consider in a discussion of the optimum parameters for magnetic hyperthermia is the influence of the size distribution on the final SAR value. To illustrate this, the SAR value for a MNP with a volume 30 % below the optimum volume was calculated in each case. The results illustrate the loss of SAR due to the size distribution of MNPs. The results are plotted as dashed lines along with the previous data.

Fig. 10 displays an essential result for magnetic hyperthermia and is worth a detailed comment beginning with the high anisotropy MNPs described by the LRT. First, the LRT shows that the maximum achievable SAR increases with a reduction in anisotropy. This is obvious from Eqns. (36)-(39): for a MNP with the optimum size, the resonating term $\frac{\omega \tau_R}{1 + \omega^2 \tau_R^2}$ is maximal and always equals $\frac{1}{2}$. Because $A \propto V$ [see Eqns. (36)-(38)] and because the optimum volume is inversely proportional to the anisotropy [see Eqn. (39)], the maximal SAR value increases with decreasing anisotropy. In this regime, the effect of the size distribution is dramatic: the SAR decreases by more than one order of magnitude for non-optimum NPs.

After passing the blank space to the left of these curves where the transition between the two regimes is out of the domain of validity of both models used here, the SAR of the optimum NPs displays a plateau: in this region, the SAR can be maximized by tuning the NP volume to adjust their coercive fields because an optimum volume satisfying Eqns. (40) and (42) always exists. With absolutely no size distribution, all of the MNPs in this anisotropy range could be perfect candidates for magnetic hyperthermia. However, it is observed that the NPs with a weak anisotropy are less sensitive to size distribution effects. At the left extremity of this plateau, the size distribution effects are cancelled. For $\phi = 0$, this occurs when $\mu_0 H_{max} \approx \mu_0 H_K$, and for the random orientation case it occurs when $0.81 \mu_0 H_{max} \approx 0.48 \mu_0 H_K$, i.e., when the targeted coercive field equals the low temperature coercive field of the material. Therefore, the relation giving the optimum anisotropy for MNPs with a known magnetization is

$$K_{opt} = C \frac{\mu_0 H_{max} M_S}{2}, \qquad (46)$$

with $C = 1$ for $\phi = 0$ and $C = 1.69$ for the random orientation case.

At the left of this plateau, a decrease of the SAR is observed. This occurs when the anisotropy field of the NP is so weak that there is no solution to Eqns. (40) and (42). In this case, the coercive field of the NPs is $\mu_0 H_K$ for $\phi = 0$ and $0.48 H_K$ for a random orientation. Then, the SAR calculated using Eqns. (25) and (26) leads to the observed decrease.

At the optimum anisotropy, the optimum volume of the MNPs diverges and tends toward infinity [see Figs 9(a) and 9(b)]. This means that, in principle, a large single-domain MNP with the optimum anisotropy would be the perfect object. However, increasing the size of the NPs too much leads to several problems:

- i) It leads to the transition toward multi-domain NPs. They might be interesting objects for magnetic hyperthermia because the coercive field value is also influenced by the size near this transition (see Fig. 8) [26]. However, there is no simple way to calculate the optimum size and the hysteresis area for such nano-objects.
- ii) Dispersing and stabilizing NPs in a colloidal solution is all the more difficult if they have a large diameter.
- iii) A larger size favors recognition by the phagocytosis system after intravenous administration.

The optimum volume of the MNPs and the optimum anisotropy thus result from a compromise between the efficiency of magnetic hyperthermia, which requires large NPs with an anisotropy given by Eqn. (46), and other factors for which a large size could be detrimental. If for any reason the volume of the MNPs needs to be limited to a maximum value, the equations used so far allow one to deduce easily what would be the optimum anisotropy for a given volume. In any case, Eqn. (46) provides the approximate target anisotropy to optimize the SAR of MNPs.

3. Optimum materials.

In Fig. 11, Eqn. (46) is plotted along with bulk parameters of several magnetic materials with no consideration of their toxicity. It is emphasized that the magnetic anisotropy in MNPs is generally larger than that of the bulk because of surface effects and/or stoichiometry problems (in alloys and oxides). In the iron oxide family, magnetite NPs displaying the bulk anisotropy would be ideal candidates. Because the hysteresis area is directly proportional to $M_S$, metallic materials with high magnetization are required to reach the highest SARs. In this case, FeCo alloys would be perfect, but their probable toxicity could be a severe problem. Iron, which is not intrinsically toxic, could be a good candidate, but presents in its crystalline form an anisotropy value too large for 20 mT applications. $Fe_{1-x}Si_x$ alloys have both a reduced anisotropy and magnetization and could represent an interesting compromise. Also, amorphous iron should display a reduced anisotropy. However, the possibility of creating MNPs with a reduced anisotropy using amorphous iron or $Fe_{1-x}Si_x$ alloys still needs to be demonstrated.

4. The Brownian motion.
a. Influence of Brownian motion on hyperthermia properties.

When MNPs are in a fluid, they can rotate physically under the influence of the magnetic fluid, similarly to a compass, until the magnetization is aligned with the magnetic field. This is known as relaxation by Brownian motion. In a standard hyperthermia experiment, the relaxation by Brownian motion and the relaxation by magnetization reversal described above are both possible, which leads to a global hysteresis loop resulting from the two mechanisms. Whether the relaxation occurs only by Brownian motion or by both mechanisms, the heating during one cycle still simply equals the hysteresis loop area $A$. The influence of Brownian motion can be easily incorporated into the LRT [10]. A Brownian relaxation time is defined as

$$\tau_B = \frac{3\eta V_H}{k_B T}, \qquad (47)$$

where $\eta$ is the viscosity of the solvent and $V_H$ is the hydrodynamic volume of the MNPs. The relaxation time $\tau_R$, which includes the Néel and Brownian relaxation times, is then defined as

$$\frac{1}{\tau_R} = \frac{1}{\tau_N} + \frac{1}{\tau_B}. \qquad (48)$$

Here again, it must be kept in mind that the LRT including Brownian motion has a restricted domain of validity, which is discussed in detail in Ref. [11] : the LRT is valid for small magnetic fields, i.e., for $\xi < x$ with $x$ depending on the value of $\frac{\omega \tau_B}{\xi}$. For any value of the parameters, numerical results have been obtained by Raiker *et al.* with the restrictive hypothesis that the relaxation by magnetization reversal is not possible [11]. The hysteresis loop of a MNP in a magnetic fluid when both magnetization reversal and Brownian motion are allowed has to our knowledge not been solved in the most general case.

### b. Optimum volume calculation

In a situation where only reversal by Brownian motion would occur, the LRT indicates that the optimum volume is the one for which $\omega \tau_B = 1$. For an applied frequency of 100 kHz and the viscosity of water, this leads to an optimum radius of around 15.5 nm independently of any magnetic parameters. However, there are some ranges of parameters (MNP magnetization, magnetic field amplitude and frequency) for which the optimum size given by this simple equation is not correct because it is out of the domain of validity of the LRT; the only way to know it is to use numerical calculations [11]. In the case of relaxation by Brownian motion, there is so far no equivalent to the simple analytical formula we used in the case of relaxation by magnetization reversal, which would be valid over a wide range of parameters. Theoretical progress is required on this point.

### c. Why Brownian motion is not the way to optimize thermal effects

Progress in the development of analytical formulas and/or the use of numerical simulations could lead to progress in the determination of the optimum parameters of MNP heating through their Brownian motion. One can imagine that eventually MNPs with such optimum magnetic properties could be synthesized. However, we think that this method is less promising than the one consisting of optimizing the SAR and using heating by magnetization reversal. The main reason is the following: the Brownian motion depends strongly on the environment and the aggregation state of the MNPs because the hydrodynamic volume of the MNP is the main parameter governing the Brownian motion. Thus, two aggregated MNPs or a MNP functionalized and linked to a tumor cell would display SAR values very different to that of a free MNP. Moreover, the application of an alternating magnetic field can lead to the formation of chains or columns of MNPs with very different Brownian properties [27, 28]. We think that the ideal objects for magnetic hyperthermia should display a SAR that is as independent as possible from such phenomena. This requires that the physical rotation of the MNP in the alternating field is blocked for all MNPs, which should thus have a large hydrodynamic diameter. This would ensure that individual free MNPs and MNPs that are a part of aggregates have the same heating properties. Coating the magnetic core using PEG or dextran layers to ensure the bioavailability and targeting of MNP is already a process tending toward this goal.

However, it is not completely impossible that MNPs moving inside a tumor under the influence of an alternating magnetic field might cause other damage to the cells than simply that resulting from the increase of temperature only.

### 5. Comparison with experimental results

In this last subsection, we will present a short summary of experimental results on a few selected materials. A comparison between theoretical maximum SARs and experimentally measured SARs will be done. A more detailed review on experiments can be found in Ref. [29].

The maximum hysteresis area $A_{max}$ that can be obtained in a hyperthermia experiment is

$$A_{max} = 4\mu_0 H_{max} M_S. \qquad (49)$$

If the hysteresis area is expressed in energy per unit of mass, which is usual in hyperthermia, $M_S$ should be replaced in this equation by the saturation magnetization per unit of mass $\sigma_S$, which leads to

$$A_{max} = 4\mu_0 H_{max} \sigma_S \qquad (50)$$

An interesting way to represent the experimentally measured area $A_{exp}$ is

$$A_{exp} = 4\alpha\mu_0 H_{max} \sigma_S \qquad (51)$$

In this formulation, $\alpha$ is a dimensionless parameter that characterizes the relative area of the hysteresis loop with respect to the ideal square. $\alpha = 1$ for a perfectly optimized system with the easy axes of all NPs aligned with the magnetic field; $\alpha = 0.39$ for an optimized system with a random orientation of the easy axes [see Eqn. (43)]. In a sense, $\alpha$ represents the degree of optimization of a given system. In Table 1, $\alpha$ is calculated from experimental results obtained on various materials of interest. For each material, the highest value in the literature was chosen. This table shows the high degree of optimization that has already been achieved in the magnetic hyperthermia properties of iron oxide nanoparticles. Because the $\alpha$ value is already nearly optimal for randomly oriented nanoparticles, further improvements will necessarily imply that the MNPs are oriented by the magnetic field during hyperthermia experiments. This table also shows that MNPs composed of high magnetization materials have not yet reach such a degree of optimization, and much higher SAR values could be reached by a better control of the nanoparticles' size, anisotropy, dispersion and magnetization values.

### IV. Conclusion

This article presents a rigorous approach to the calculation of the hysteresis area of a single-domain in the macrospin approximation, including a detailed study of the validity of the analytical expressions. The conclusions about the applications of such nanoparticles for magnetic hyperthermia are clear and simple in this framework. We hope these conclusions will guide experimentalists both in the synthesis of high-quality materials and in the analysis of experimental data. An important conclusion of this study is that the anisotropy of the synthesized MNPs is a key parameter to understand and tune the magnetic hyperthermia properties. Often neglected, it should, on the contrary, become central in experimental articles on magnetic hyperthermia.

However, it should not be overlooked that several important points have not been treated in this article and have a strong influence for the required application. i) At room temperature and high frequency, what is the domain of validity of the macrospin approximation and thus of the SWMBTs used to predict hyperthermia properties? ii) To what extent can the single-domain/multi-domain transition be predicted and used to tune the coercive fields of magnetic nanoparticles to maximize hyperthermia? Is the shape and area of the hysteresis loop obtained in this case interesting for magnetic hyperthermia applications? iii) How are the results for the optimum parameters modified when taking into account the magnetic interactions, which are known to have a deep influence on magnetic properties both in the superparamagnetic and the ferromagnetic regimes? We hope that these questions will stimulate theoretical work on this subject.

**Acknowledgements :**
We acknowledge L.-M. Lacroix, S. Lachaize and Y. Raikher for fruitful discussions. This work was supported by the InNaBioSanté Foundation.

**Table :**

| Materials | $\sigma_S$ (300 K) (Am$^2$/kg) | $\mu_0H_{max}$ (mT) | $A_{max}$ (mJ/g) | $A_{exp}$ (mJ/g) | $\alpha$ |
|---|---|---|---|---|---|
| FeO$_x$ MNPs | 100 | 13.8 | 5 [30] | 1.5 | 0.3 |
| magnetosomes (a) | 100 | 12.5 | 5 [31] | 1.3 | 0.26 |
| magnetosomes (b) | 100 | 12.5 | 5 [31] | 2.3 | 0.46 |
| Co MNPs | 162 | 31.2 | 20.6 [32] | 3.25 | 0.16 |
| FeCo MNPs | 240 | 29 | 27.8 [19] | 1.5 | 0.054 |
| Fe MNPs | 218 | 66 | 57.5 [33] | 5.6 | 0.097 |
| CoFe$_2$O$_4$ | 75 | 31.1 | 9.35 [34] | 0.63 | 0.067 |

Table I: Summary of experimental results on various materials of interest. The "$\sigma_S$" column gives the bulk magnetization per unit mass at 300 K. "$\mu_0H_{max}$" is the magnetic field at which the experiments were conducted. "$A_{max}$" gives the theoretical maximum hysteresis area that could have been measured, which was calculated using Eqn. (50). "$A_{exp}$" gives the hysteresis area experimentally measured in these conditions. "$\alpha$" is calculated using Eqn. (51). The rows labeled "magnetosomes" correspond to iron oxide NPs synthesized by bacteria. NPs were (a) randomly oriented or (b) aligned with the magnetic field.

**References :**

*Electronic mail : julian.carrey@insa-toulouse.fr

**Figures :**

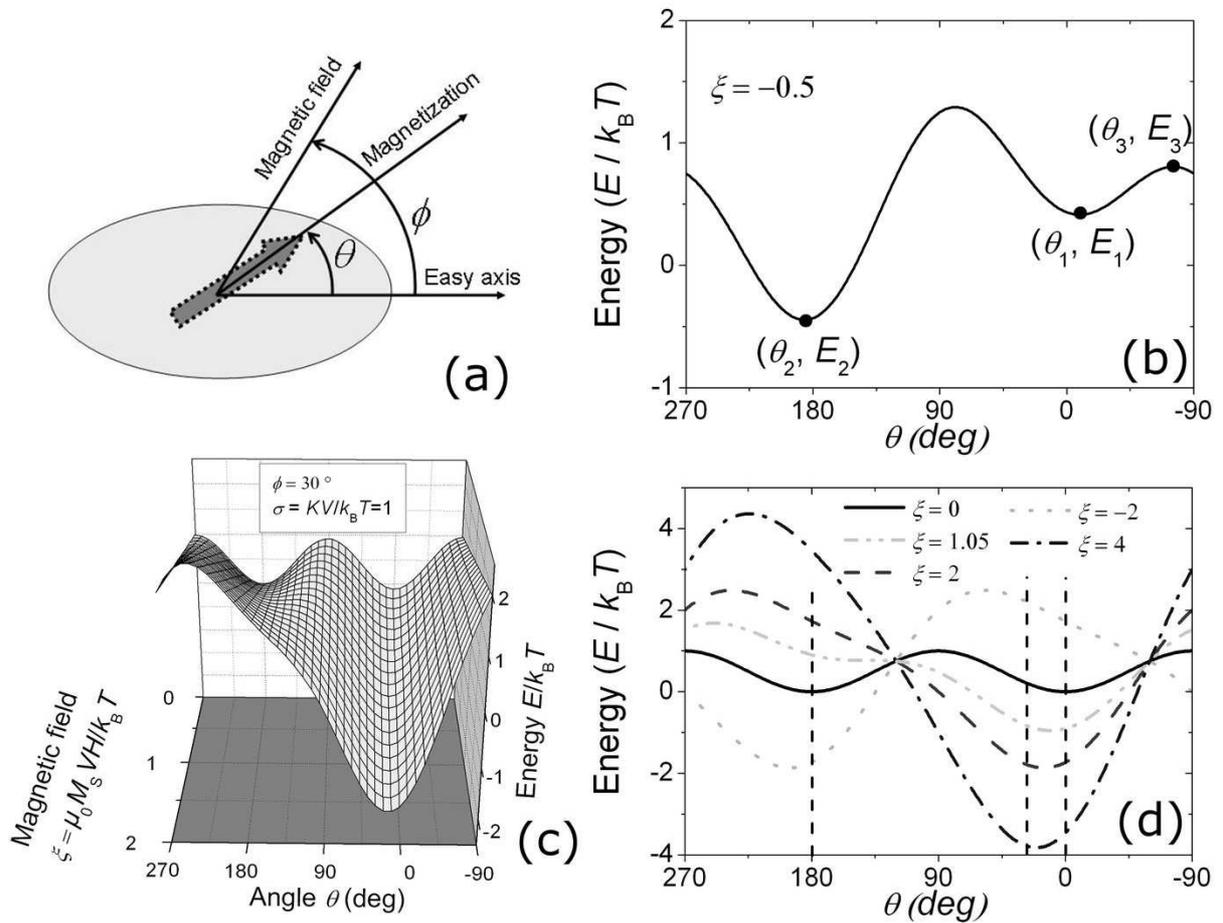

Figure 1 : (a) Schematic representation of a uniaxial single-domain MNP. The large arrow represents the magnetization. (b) Illustration of the three extrema of the energy landscape with $\xi = -0.5$, $\phi = 30°$ and $\sigma = 1$. (c) The energy of a NP as a function of $\theta$ is plotted as a function of $\xi$ for $\phi = 30°$ and $\sigma = 1$. (d) The energy of a NP as a function of $\theta$ for $\xi = -2, 0, 1.05, 2$ and $4$. In this example, the energy barrier between the two minima disappears for $\xi \approx 1.05$. Vertical dashed lines are drawn at $\theta = 0, 30$ and $180°$.

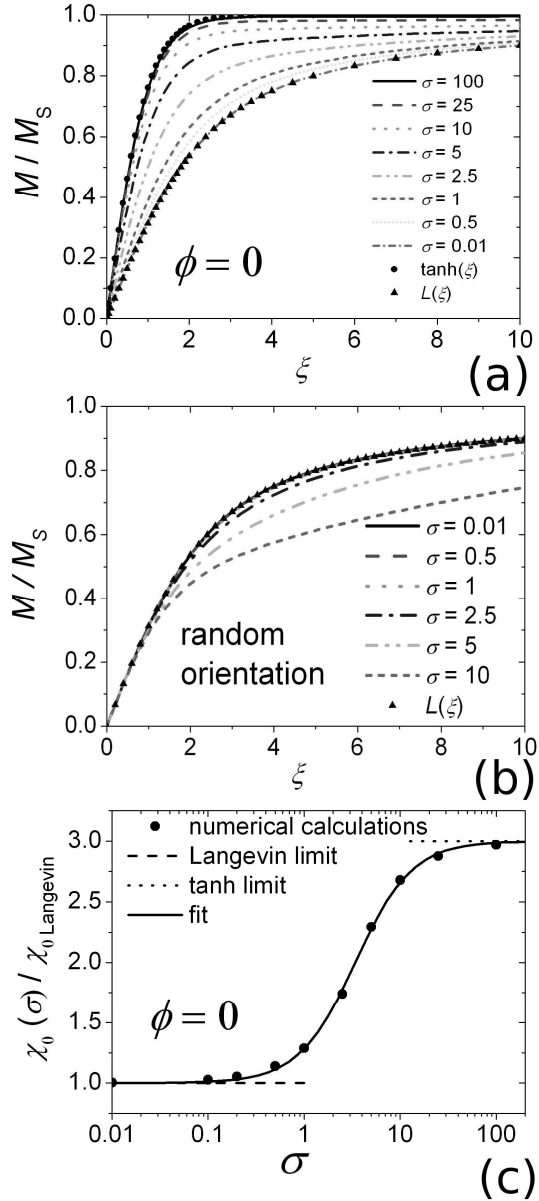

Figure 2 : Results of the equilibrium functions. (a) and (b): Numerical calculations of the hysteresis loop for MNPs at thermal equilibrium for various values of their reduced anisotropy $\sigma$. The $L(\xi)$ function and the $\tanh(\xi)$ function are plotted alongside the data for comparison. (a) The easy axes are aligned with the magnetic field ($\phi = 0$). (b) The easy axes are randomly oriented in space. (c) Evolution of the initial slope of the hysteresis loop as a function of $\sigma$ when $\phi = 0$. The dots are extracted from the numerical simulations. The line is a phenomenological fit using Eqn. (12). The dashed and dotted lines show the initial slope of the $L(\xi)$ and $\tanh(\xi)$ functions, respectively.

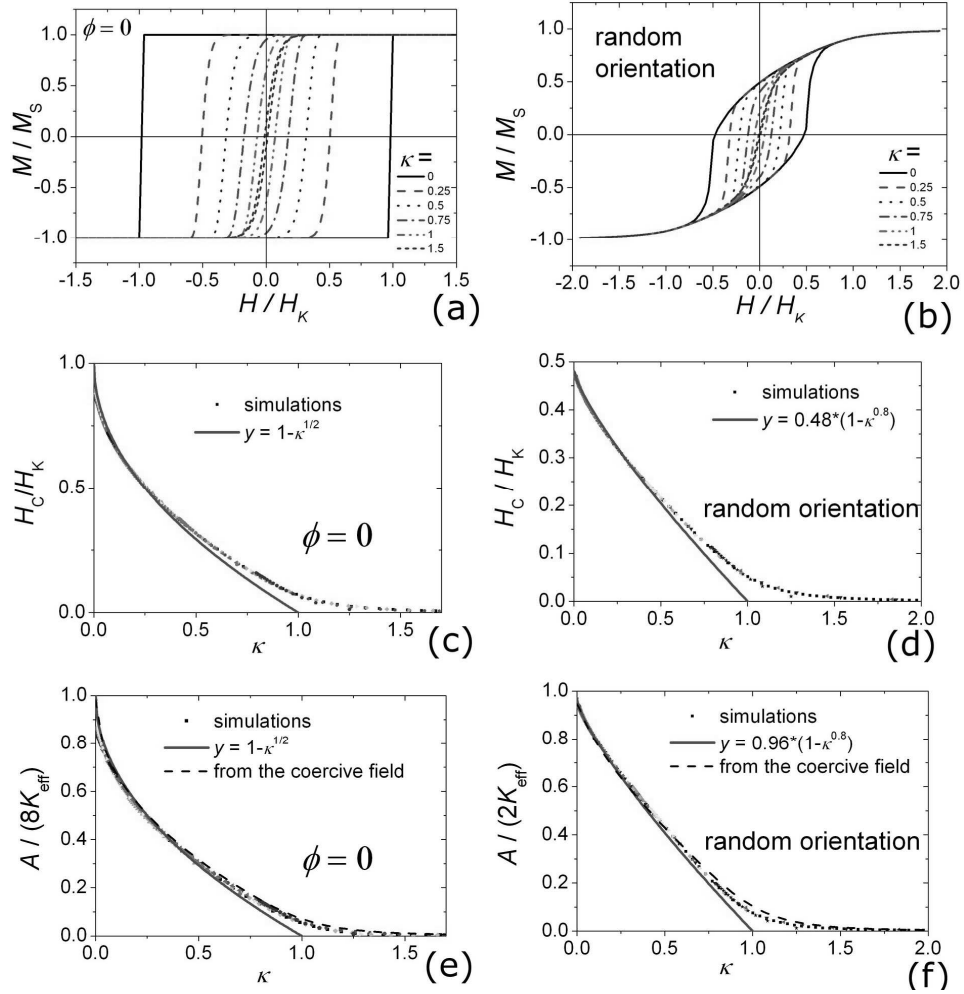

Figure 3 : Results on SWMBTs. (a) and (b) Examples of major hysteresis loops for values of $\kappa$ ranging from 0 to 1.5. (c) and (d) Normalized coercive field as a function of $\kappa$. The dots corresponds to simulation results with $H_{\max}$, $K$, $V$, $T$ and $f$ varying over a wide range of experimentally relevant values. Solid lines correspond to Eqns. (22) or (24). The dotted line is the area calculated from the true coercive field of the hysteresis loops using Eqn. (25) or (26). (e) and (f) Normalized hysteresis area as a function of $\kappa$. Solid lines are calculated by combining Eqn. (22) with Eqn. (25) or Eqn. (24) with Eqn. (26). The dashed line is the area calculated from the true coercive field of the hysteresis loops using Eqn. (25) or (26). (a), (c) and (e): $\phi = 0$. (b), (d) and (f): random orientation.

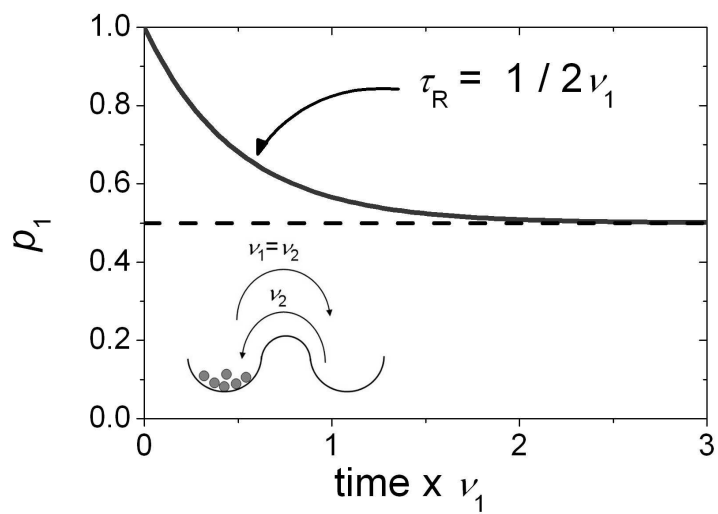

Figure 4 : Evolution of the probability $p_1$ to find a nanoparticle in the initially full potential well as a function of $t \times \nu_1$. The result is an exponential decay function with a time constant of $1/2\nu_1$. (inset) Illustration of the emptying of an initially full potential well through a reversible jump over the barrier.

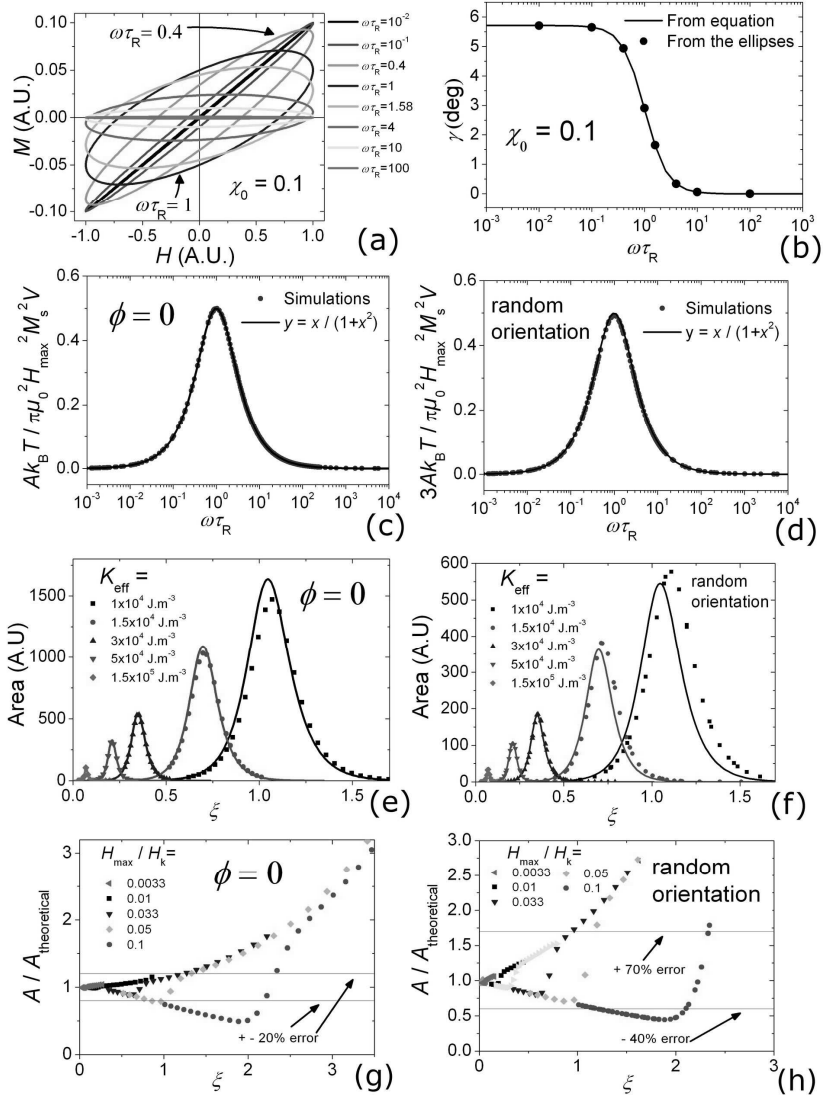

Figure 5: Results of the LRT. (a) Evolution of the hysteresis loop as a function of $\omega\tau_R$ is plotted using Eqns. (30), (31), (32) and (33) for $\chi_0 = 0.1$ and $H_{max} = 1$. (b) Angle between the long axis of the ellipse and the abscise axis $\gamma$ as a function of $\omega\tau_R$. The line is deduced from Eqn. (35); the dots are a direct measurement from the ellipses plotted in (a). (c) and (d) Normalized hysteresis areas as a function of $\omega\tau_R$. Each dot corresponds to a numerical simulation and the line to LRT [Eqns. (36) and (37)]; values of $K_{eff}$, $V$, $f$ and $\mu_0 H_{max}$ were varied over a wide range of parameters keeping $\xi$ and $\mu_0 H_{max} / H_K$ well below 1. (e) and (f) Hysteresis areas obtained by simulations (dots) are plotted along the theoretical areas provided by Eqn. (36) or (37) (lines) as a function of $\xi$ for various values of $K_{eff}$. $\mu_0 H_{max} = 1$ mT, $f = 100$ kHz, $T = 300$ K and $M_S = 10^6$ Am$^{-1}$. (g) and (h) Data similar to the previous ones except that the hysteresis area is divided by the theoretical area and that data for higher values of $\xi$ are shown. The corresponding values of $\mu_0 H_{max} / H_K$ are provided. The horizontal dashed line illustrates the discrepancy between Eqns. (36) or (37) and the simulations when $\xi > 1$. (c), (e) and (g): $\phi = 0$. (d), (f) and (h): random orientation.

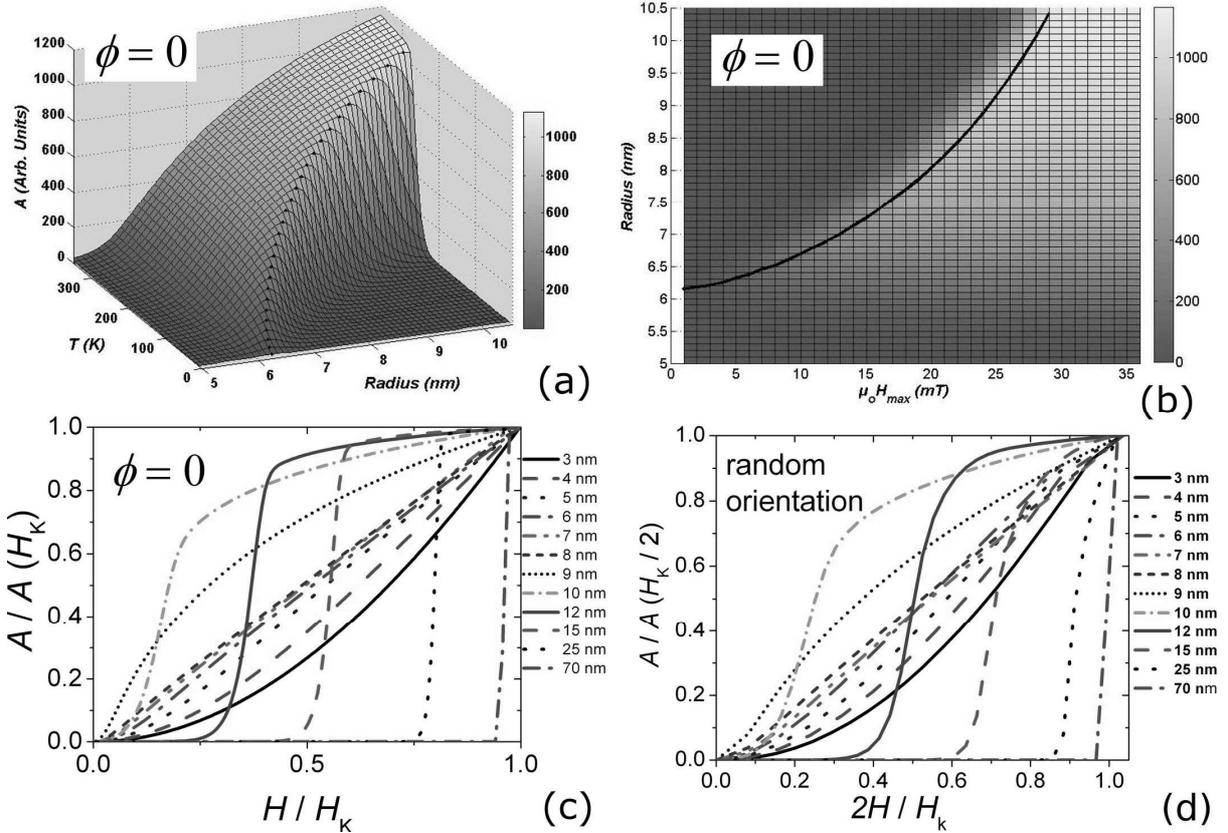

Figure 6 : Numerical simulations of the hysteresis area. When they are not varied, the parameter values are $K_{eff} = 13000$ J.m$^{-3}$, $M_S = 10^6$ A.m$^{-1}$, $f = 100$ kHz, $\mu_0 H_{max} = 20$ mT, $T = 300$ K and $\nu_0^1 = 10^{10}$ Hz. (a) Evolution of area as a function of the radius and temperature. The dots represent the maximum area for a given temperature. (b) Evolution of area as a function of the magnetic field and the radius. The line represents the maximum area at a given magnetic field. (c) and (d) Normalized hysteresis area as a function of the normalized magnetic field for various MNP radii. (a), (b) and (c) $\phi = 0$. (d) random orientation.

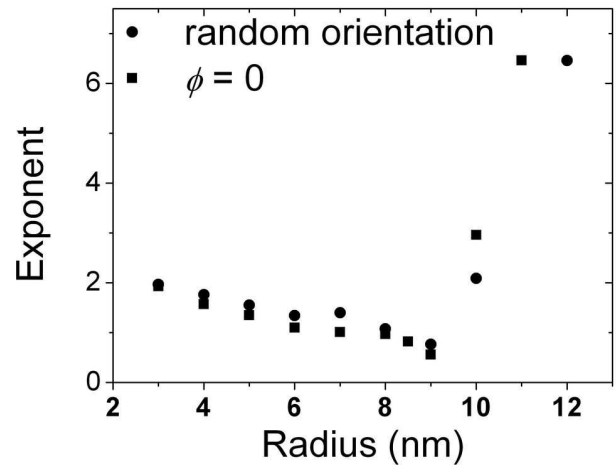

Figure 7 : Exponent of the best power law fit of the curves shown in Fig. 6(c) and (d). When there was an inflexion point in the curve, the fit was performed only up to this point.

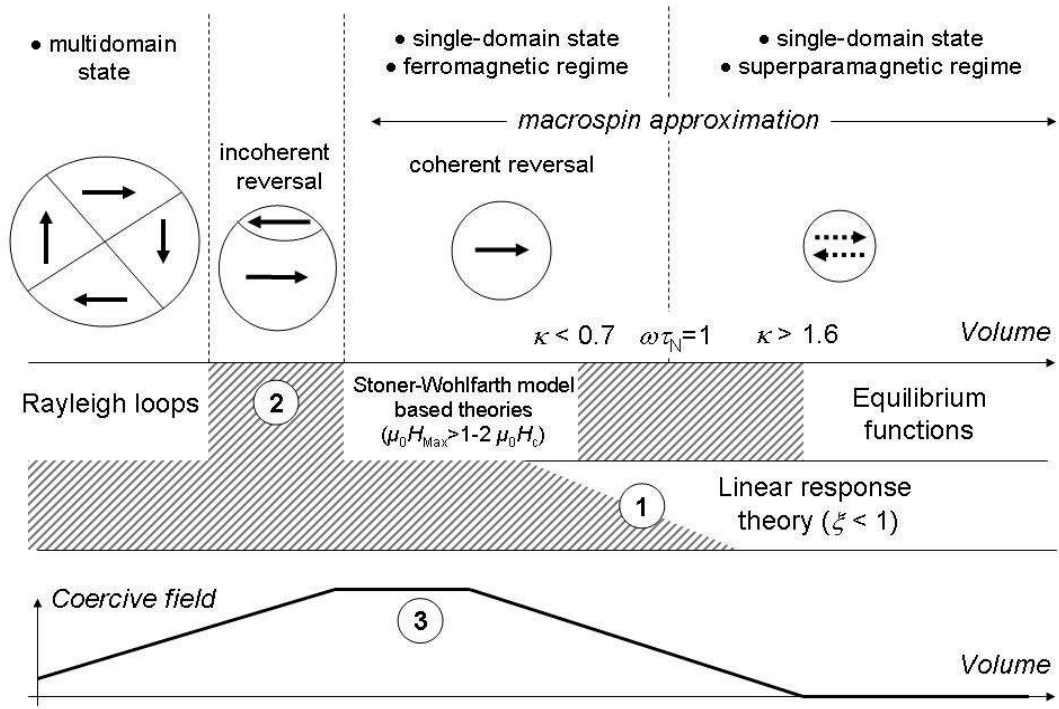

Figure 8 : Schematic representation of the evolution of the magnetic properties of MNPs as a function of their volume and of the models suitable to describe them. The label (1) illustrates that the maximum magnetic field for which the LRT is valid decreases with increasing volume. The label (2) is the domain where incoherent reversal modes occur so SWMBTs are not valid anymore. The label (3) shows a plateau in the volume dependence of the coercive field.

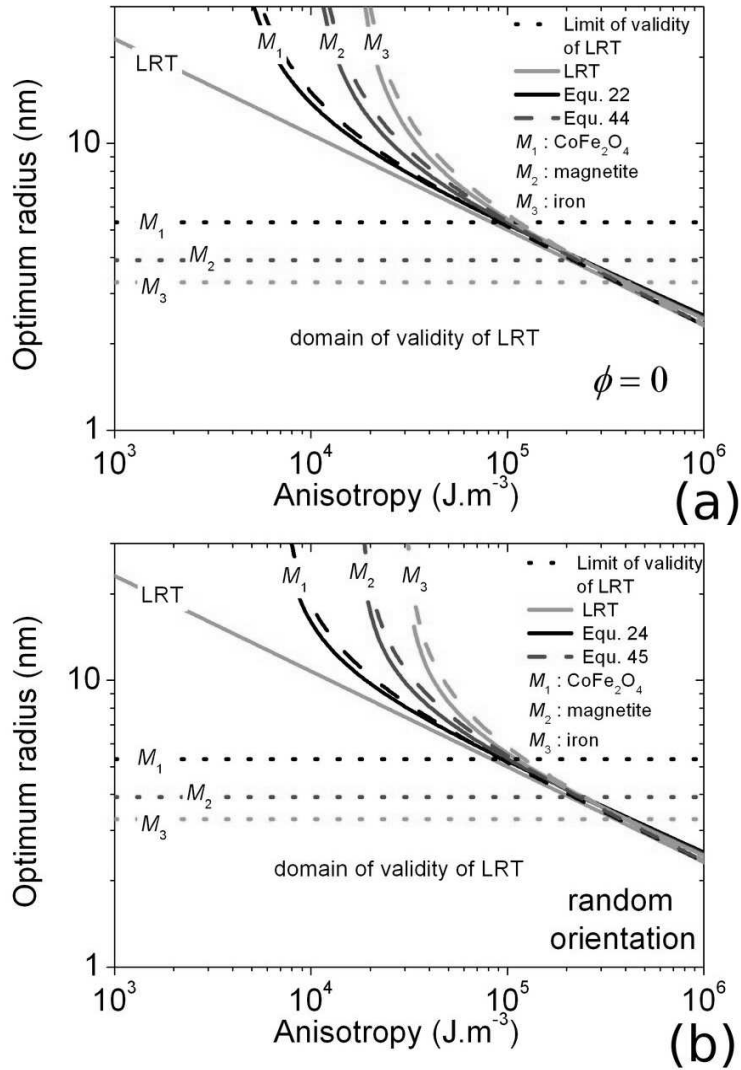

Figure 9 : Calculations of the optimum radius for MNPs as a function of their anisotropy using the LRT and SWMBTs for three different values of $M_S$ labeled as $M_1$, $M_2$ and $M_3$. $M_1 = 0.4 \times 10^6$ A.m$^{-1}$, $M_2 = 10^6$ A.m$^{-1}$, $M_3 = 1.7 \times 10^6$ A.m$^{-1}$, $f = 100$ kHz, $\mu_0 H_{max} = 20$ mT, $T = 300$ K and $\nu_0^1 = 10^{10}$ Hz. The horizontal dotted lines show the limit above which the LRT is not valid anymore ($\xi > 1$). The LRT result is common to all graphs and is given by Eqn. (39). (a) and (b) Comparison between different formulas to calculate the optimum size. The solid lines refer to the numerical solution of Eqns. (22) and (24). The dashed line refers to Eqns. (44) and (45). (a): $\phi = 0$. (b): random orientation.

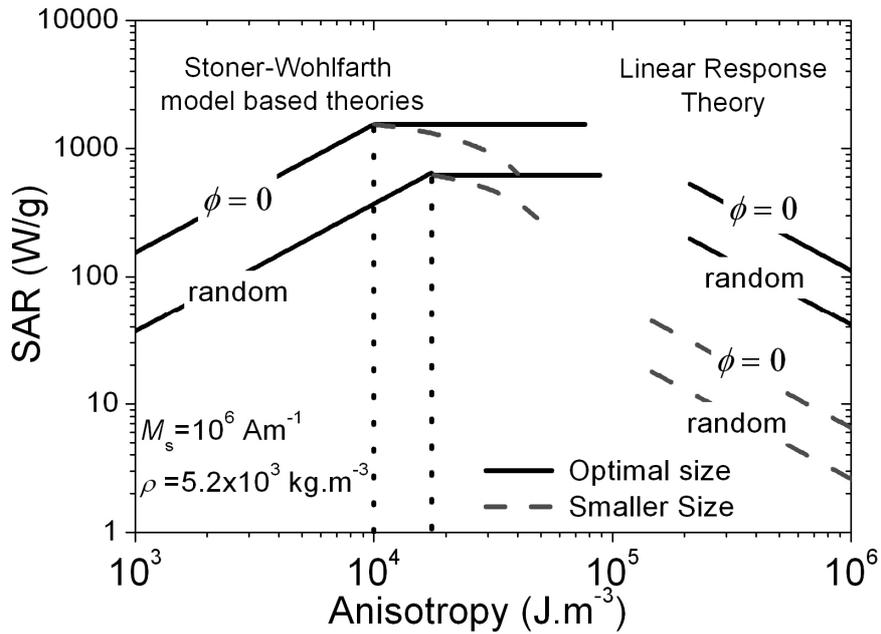

Figure 10 : Calculation of the SAR as a function of the anisotropy for NPs with the optimum volume (solid lines) and for NPs with a volume equal to 70% of the optimum volume (dashed lines). The parameters are identical to those used in the previous figure with $M_S = M_2 = 10^6$ Am$^{-1}$ and $\rho = 5.2\times10^3$ kg.m$^{-3}$. For strongly anisotropic NPs, the LRT results are provided using Eqns. (36), (38) and (39). For weakly anisotropic NPs, SWMBTs results are provided using Eqns. (19), (21), (25) and (40)-(45). The vertical dotted lines show the optimum anisotropy for this set of parameters.

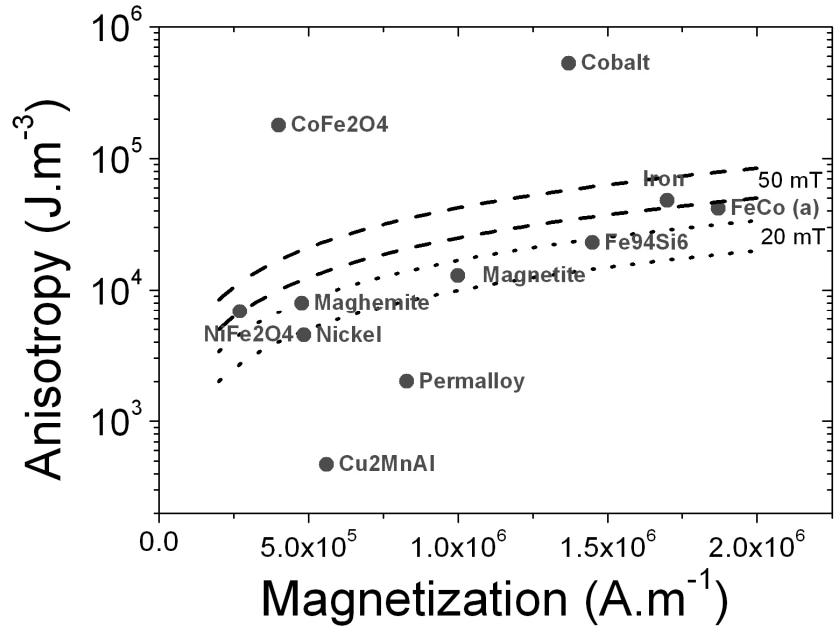

Figure 11 : Comparison between theoretical optimum parameters and experimental bulk parameters of several magnetic materials. Eqn. (46) is plotted for $\mu_0 H_{max} = 20$ mT (dotted lines) and 50 mT (dashed lines). In each case, the upper curve represents the random orientation case, and the lower one represents the $\phi = 0$ case. Label (a): for the FeCo alloy, the anisotropy of 15 nm MNPs estimated in Ref. [19] is provided.

.